\documentclass[useAMS,fleqn,usenatbib]{mnras}

\usepackage{graphicx}

\usepackage[T1]{fontenc}
\usepackage{ae,aecompl}
\usepackage{amsmath}	
\usepackage{amssymb}	
\usepackage{xcolor}
\usepackage{hyperref}

\title[Large-scale magnetic field in accretion discs]{Large-scale magnetic field in the accretion discs of young stars: the influence of magnetic diffusion, buoyancy and Hall effect}
\author[S.~A.~Khaibrakhmanov, A.~E.~Dudorov, S.~Yu.~Parfenov and A.~M.~Sobolev]{S.~A.~ Khaibrakhmanov$^{1, 2}$\thanks{E-mail: khaibrakhmanov@csu.ru (SAKh)}, A.~E.~Dudorov$^{2}$\thanks{E-mail: dudorov@csu.ru (AED)}, 
S.~Yu.~Parfenov$^{1}$\thanks{E-mail: sergey.parfenov@urfu.ru (SYuP)},
and A.~M.~Sobolev$^{1}$\thanks{E-mail: andrej.sobolev@urfu.ru (AMS)}\\
$^{1}$ Ural Federal University, 51 Lenin str., Ekaterinburg 620000, Russia\\
$^{2}$ Chelyabinsk state university, 129 Br. Kashirinykh str., Chelyabinsk 454001, Russia}
\begin{document}

\date{}

\pagerange{\pageref{firstpage}--\pageref{lastpage}} \pubyear{2016}

\maketitle

\label{firstpage}

\begin{abstract}
We investigate the fossil magnetic field in the accretion and protoplanetary discs using the Shakura and Sunyaev approach. The distinguishing feature of this study is the accurate solution of the ionization balance equations and the induction equation with Ohmic diffusion, magnetic ambipolar diffusion, buoyancy and the Hall effect. We consider the ionization by cosmic rays, X-rays and radionuclides, radiative recombinations, recombinations onto dust grains, and also thermal ionization. The buoyancy appears as the additional mechanism of magnetic flux escape in the steady-state solution of the induction equation. Calculations show that Ohmic diffusion and magnetic ambipolar diffusion constraint the generation of the magnetic field inside the `dead' zones. The magnetic field in these regions is quasi-vertical. The buoyancy constraints the toroidal magnetic field strength close to the disc inner edge. As a result, the toroidal and vertical magnetic fields become comparable. The Hall effect is important in the regions close to the borders of the `dead' zones because electrons are magnetized there. The magnetic field in these regions is quasi-radial. We calculate the magnetic field strength and geometry for the discs with accretion rates $(10^{-8}-10^{-6})\,\rm{M}_{\odot}\,\rm{yr}^{-1}$. The fossil magnetic field geometry does not change significantly during the disc evolution while the accretion rate decreases. We construct the synthetic maps of dust emission polarized due to the dust grain alignment by the magnetic field. In the polarization maps, the `dead' zones appear as the regions with the reduced values of polarization degree in comparison to those in the adjacent regions.

\end{abstract}


\section{Introduction}

Contemporary star formation takes place in the magnetic rotating cores of molecular clouds which we call protostellar clouds. The collapse of protostellar clouds leads to the formation of the protostars with protostellar discs. The discs form under the action of the electromagnetic and centrifugal forces. The protostellar discs are geometrically thick self-gravitating discs. After the formation of a young star, the protostellar disc transforms into the accretion disc, where matter accretes onto the star with the accretion rate $\dot{M}=(10^{-6}-10^{-8})\,\rm{M}_{\odot}\,\rm{yr}^{-1}$ in the case of T~Tauri stars. The accretion discs of the T~Tauri stars are geometrically thin structures with masses $\sim(0.001-0.1)\,\rm{M}_{\odot}$ and sizes $(100-1000)$~au (see review of \citet{williams11}). During evolution, the accretion discs transform to the protoplanetary discs similar to the protosolar nebula.

There are some observational data concerning the magnetic fields in the accretion discs of young stellar objects. The measurements of the remnant magnetization of the meteorites indicate that the magnetic field strength in the protosolar nebula was $\sim(0.1-1)$ Gs~\citep{levy78, fu14}. \citet{donati05} reported about the measurement of the magnetic field strength of about 1~kGs in the inner regions of the FU~Ori system using the Zeeman splitting technique. \citet{girart06} made the measurements of the polarized dust emission from the low-mass protostellar system NGC~1333~IRAS~4A and found that the geometry of the magnetic field is `hour-glass' in this system. \cite{Rao_etal2014} detected linearly polarized 878 $\mu\rm{m}$ dust emission in the circumstellar disc around the IRAS~16293-2422~B protostar from the Submillimetre Array observations. The measurements indicate that the geometry of the magnetic field is complex in this system. \citet{stephens14} performed resolved measurements of the polarized 1.25~mm continuum emission from the HL~Tau disc using the Combined Array for Millimetre-wave Astronomy (\textsc{CARMA}). They suggested that the geometry of the magnetic field in the disc is complex with both toroidal and poloidal components. \citet{scox15} performed observations of the dust polarization from the circumstellar disc around the Class 0 protostar L1527 using the \textsc{CARMA}. They concluded that the geometry of the magnetic field is toroidal.

Interpretation of (sub)millimetre polarization observations of the discs around young stellar objects can be complicated. The polarization of the disc dust emission is usually interpreted as a result of the alignment of non-spherical dust grains with magnetic field. But as it was demonstrated by \citet{Yang_etal2016} the (sub)mm continumm polarization could also be a result of scattering and disc inclination. Direct measurements of the magnetic field strength based on the Zeeman effect are yet not possible and the polarization observations are not able to detect the large-scale magnetic field geometry in details. Therefore, theoretical investigations of the magnetic field in young stellar objects are needed to explain the observations and to predict what magnetic field will be detected in future high-resolution measurements.

There are two main conceptions of the origin of the magnetic field in young stellar objects: the dynamo theory and the theory of the fossil magnetic field. The latter is based on the numerical simulations and observations of the star formation in molecular cloud cores (see recent review by \citet{fmft}). One-and-half-dimensional magneto-gas-dynamic simulations of \citet{dudorov87} have shown that the initial magnetic flux of protostellar clouds is conserved partially during collapse and accretion.  These conclusions have been confirmed by a number of multi-dimensional simulations (see review by \citet{i12}). Therefore, from the point of view of the discs formation theory it is natural to assume that the magnetic field of stars and their accretion discs is the {\it fossil} one, i.e. it is the remnant of the protostellar cloud's magnetic field. The strength and geometry of the magnetic field in the accretion discs is influenced by accretion, differential rotation, magnetic diffusion and different instabilities.

Ohmic diffusion (OD) and magnetic ambipolar diffusion (MAD) are the main processes that constraint the generation of the magnetic field during the star and disc formation. OD of the magnetic field is caused by currents dissipation in gas with finite conductivity (see \citet{parker_book}). MAD is the process of plasma drift through the neutral gas under the action of the electromagnetic force. \citet{mestel56} first pointed that MAD will allow a protostellar cloud to contract across the field lines if the cloud is dense enough and the frictional coupling between plasma and neutral gas is small. 

The magnetic diffusion efficiency depends on the ionization fraction. \citet{hayashi81} estimated the conductivity of the protosolar nebula with the magnetic field taking into account ionization by cosmic rays and decay of radioactive elements. Hayashi concluded that OD prevents generation of the strong magnetic field in the regions of the terrestrial planet formation. \citet{lubow94} and \citet{agapitou96} have shown that significant inward dragging of the large-scale magnetic field lines occurs if the magnetic diffusivity is much less than the turbulent viscosity in the disc. \citet{rr96} pointed out that toroidal magnetic field can be generated in a rotating disc when magnetic diffusion is weak. In these papers, OD efficiency was described with the help of non-dimensional magnetic Prandtl number determined as the relation of the magnetic diffusivity to the turbulent viscosity.

\citet{gammie96} introduced layered accretion scenario for the T~Tauri accretion discs. He pointed out that the ionization fraction can be very low, $x<10^{-13}$, near the mid-plane of the disc due to the attenuation of cosmic rays. Magneto-rotational instability \citep[MRI, ][]{velikhov59, chandra60, bh91} is not developed and magneto-hydrodynamic (MHD) turbulence is weakened in such regions of low ionization fraction. He called these regions `dead' zones. The accretion takes place only in the surface layers of the discs in this model. \citet{sano00} calculated the ionization fraction of the minimum mass solar nebula \citep[MMSN, ][]{cameron73, mmsn}. The `dead' zone was determined as the region where wavelength of most unstable MRI mode exceeds the accretion disc scale height. This critical wavelength depends on the magnetic field strength and the ohmic diffusivity. \citet{sano00} investigated dependence of  `dead' zone size on the plasma parameter $\beta=(100,\,1000)$ and on the dust grain size. They have shown that the `dead' zone shrinks as the dust grains radius increases and $\beta$ decreases. The `dead' zones are located at the distances from 1~au to (10-20)~au inside the disc.

\citet{mohanty13} calculated the characteristics of the `dead' zones in the frame of the MMSN model. The `dead' zones were treated as the regions where MRI growth rate is less than the dissipation rate. The authors considered damping of MRI by both OD and MAD. They concluded that MAD determines the outer boundary of the `dead' zone. Using local shearing-box simulations of the MMSN discs, \citet{bai13} has shown that MRI is completely suppressed in the region from $0.5$~au to $(5-10)$~au near the mid-plane due to OD and in the surface layers due to MAD. 

Global MHD simulations of accretion discs dynamics were performed in the ideal MHD approach \citep{fromang06, flock11, suzuki14} and in the resistive limit \citep{dzyu10}. \citet{gressel15} carried out global simulations of the protoplanetary discs using the \textsc{nirvana} III code. The calculations were done for the part of the disc between $1$~au and $5$~au in two dimensions taking into account OD and MAD and ionization by FUV and X-rays only.

Buoyancy is another mechanism that influences the magnetic field in the accretion discs. \citet{parker_book} has shown that the magnetized plasma in gravitationally stratified fluid is buoyantly unstable. The magnetic field splits into flux tubes rising from the fluid because of the buoyancy force. Various aspects of the magnetic buoyancy in the accretion discs have been investigated, such as the dynamics of slender magnetic flux tubes \citep{sakimoto89, schram93, ziegler01}, the generation of turbulence \citep{rozyczka96}, the formation of hot corona and bursts activity \citep{miller00, machida00}, magnetic dynamo \citep{tout92, johansen08}.   \cite{zhilkin12} performed three-dimensional MHD simulations of flows in the contact binary systems and have shown that buoyancy can constraint the generation of the toroidal magnetic field.The influence of the buoyancy on the fossil magnetic field strength has not been investigated yet.

Conductivity is anisotropic in the magnetized plasma \citep{alfven_book, cowling_book}. Electrons drift causes the Hall current directed perpendicular to the magnetic field lines. \citet{urpin91}, \citet{shalybkov97} have shown that the Hall current can generate the toroidal field from the poloidal one in the conducting medium. \citet{vainshtein00} also have found that the Hall currents can lead to the exchange of energy between different components of the magnetic field in the stratified plasma.

\citet{wardle99a} calculated the conductivity tensor for the molecular gas taking into account the Hall effect. They have found that the Hall effect contribution to the conductivity is important for the densities between $10^7$ to $10^{11}\,\rm{cm}^{-3}$ in the case of Mathis-Rumpl-Nordsieck (MRN) grain-size distribution. Analytical and numerical investigations have shown that the Hall effect can increase or decrease MRI growth rate \citep{wardle99b, balbus01, sano02} and `dead' zone size \citep{wardle12} depending on the direction of the magnetic field vector ${\bf B}$ with respect to the rotation axis direction ${\bf \Omega}$. \citet{lesur14} and \citet{bai14} performed local shearing-box simulations of the protoplanetary discs dynamics taking into account OD, MAD, and the Hall effect. They analysed the role of these effects at fixed radial distances from a star. Simulations have shown the artificial generation of the strong azimuthal magnetic field in the midplane of the disc in the case $\left({\bf B}\cdot{\bf \Omega}\right)>0$. This effect is caused by the symmetry features of the shearing-box approximation and artificial limitation of the diffusivity.

Thus, the role of OD and MAD has been mainly investigated in application to MRI and `dead' zones characteristics.  \citet{guilet14} developed the approach proposed by \citet{lubow94}. To investigate the poloidal magnetic field dragging in the accretion discs, they used the vertically averaged values of the advection and diffusion rates which take into account the back-reaction of the mean magnetic field on the flow. The effective turbulent diffusivity was characterized by \citet{ss73} $\alpha$ parameter provided that the magnetic Prandtl number equals 1. This approach allowed to obtain more efficient dragging of the poloidal magnetic field than in previous works. Assuming  balance between the advection and diffusion of the magnetic field, \citet{takeuchi14a} derived the steady-state radial distribution of the vertical magnetic field $B_z(r)\propto r^{-2}$ reflecting the magnetic flux conservation law. Using this profile, the authors obtained the upper limit on the magnetic field strength, $0.1$~Gs at $r=1$~au and $\sim1$~mGs at $10$~au. \citet{takeuchi14b} investigated the relaxation of the poloidal magnetic field to this steady state taking into account Ohmic diffusion only. Both \citet{guilet14} and \citet{takeuchi14a} did not consider the toroidal magnetic field in the disc. \citet{fmfadys} (DK14, hereafter) developed the MHD model of the stationary geometrically thin low-massive accretion discs with the fossil magnetic field. The model for the accretion disc contains \citet{ss73} equations, the induction equation with OD and MAD, the ionization balance equations taking into account thermal ionization, shock ionization by cosmic rays, X-rays and radioactive elements, radiative recombinations, recombinations on various dust grains. The dust particles are considered to be well-mixed with the gas.  DK14 concluded that OD and MAD constraint the generation of the magnetic field inside the `dead' zones, and the magnetic field there is quasi-poloidal. The magnetic field can be quasi-azimuthal or quasi-radial in the outer regions depending on the ionization rates and dust characteristics.

Many previous works a priori determined the magnetic field strength and/or magnetic diffusion efficiency in a variety of ways, in particular assuming constant plasma parameter $\beta$ and/or constant magnetic Prandtl number, etc. Following DK14, we calculate the magnetic field strength and geometry on the basis of the theory of the fossil magnetic field, i.e. using the initial conditions of the disc formation. Currently, 2D and 3D MHD simulations of the dynamics of protostars with accretion discs are difficult because of the limitations on the temporal and spatial resolution. Our approach is semi-analytical. We do not perform full 3D simulations, but calculate all three components of the magnetic field taking into account many physical effects, such as ionization, recombinations, magnetic diffusion, and thermal effects.

In this paper, we modify our basic accretion disc model in order to take into account the magnetic buoyancy and the Hall  effect. In section \ref{Sec:BaseModel}, we describe the model of the accretion disc. Details can be found in the paper DK14. The modification of the model is carried out in section \ref{Sec:Modif}. We derive the induction equation with OD, MAD, buoyancy and Hall effect in section \ref{Sec:IndEq}. Estimation of the buoyancy velocity is given in section \ref{Sec:Vb}. The results of calculation of the magnetic field are presented in section \ref{Sec:Results}. We investigate influence of the buoyancy and the Hall effect on the fossil magnetic field in sections \ref{Sec:ResBu} and \ref{Sec:ResHall}, respectively. In section \ref{Sec:Mdot}, we study the intensity and geometry of the fossil magnetic field with different mass accretion rates. To demonstrate how the disc magnetic field can affect the (sub)mm continuum polarization, we performed radiative transfer calculations. First, we calculated the dust temperature, $T_{\rm{d}}$, distribution in the disc (Section~\ref{sec:hyperion}). This distribution was then used to calculate the intensity and polarization of the (sub)mm continuum emission (Section~\ref{sec:lime}). We summarize our findings in section \ref{Sec:Summary}.

\section{MHD model of the accretion disc}
\label{Sec:BaseModel}

\subsection{Basic model}
We consider the geometrically thin axially symmetric accretion disc of the star with mass $M$. The self-gravity of the disc is neglected comparing to the gravity of the star. In cylindrical coordinate system $\{r, \, \varphi,\, z\}$, the gas velocity and fossil magnetic field have components ${\bf V} = \{V_r,\, V_{\varphi},\, V_z\}$ and ${\bf B} = \{B_r,\, B_{\varphi},\, B_z\}$ . 

The basic equations of the model are magneto-gas-dynamic equations with OD and MAD (see DK14). We neglect the electromagnetic force in the equations. In this case, the radial structure of the disc is described by equations of \citet{ss73}:
\begin{eqnarray}
\dot{M}\Omega_k f &=& 2\pi\alpha\Sigma V_s^2,\label{Eq:AngMom}\\
\dot{M} &=& -2\pi rV_r\Sigma,\label{Eq:Mdot}\\
\frac{4\sigma_{\rm{sb}}T^4}{3\kappa\Sigma} &=& \frac{3}{8\pi}\dot{M}\Omega_k^2f,\label{Eq:Teff}\\
H &=& \frac{V_s}{\Omega_k},\label{Eq:H}
\end{eqnarray}
where $\dot{M}$ is the mass accretion rate, 
\begin{equation}
\Omega_k = \sqrt{\frac{GM}{r^3}}\label{Eq:Omega_k}
\end{equation}
-- the Keplerian angular velocity, $f = \left(1 - \sqrt{r_{\rm{in}} / r}\right)$, $r_{\rm{in}}$~--~the radius of the disc inner edge, $\alpha$~--~turbulence parameter, $\Sigma$~--~gas column density, $V_s=\sqrt{R_gT/\mu}$~--~isothermal sound speed, $T$~--~gas temperature, $H$~--~accretion disc scale height, $\mu=2.3$~--~the mean molecular weight of the gas.

The opacity coefficient $\kappa$ is defined as the power-law function of the gas density $\rho$ and temperature $\kappa=\kappa_0\rho^aT^b$, where constants $\kappa_0$, $a$ and $b$ are evaluated according to \cite{semenov03}. 

The inner boundary of the accretion disc is determined by the magnetosphere radius of the star. The star is considered to have dipole magnetic field with surface magnetic field $B_{\rm{s}}$. The outer boundary of the disc corresponds to the distance, where the pressure is equal to the pressure of molecular cloud cores with typical density. 

In order to calculate the magnetic field geometry, we need to specify the vertical structure of the disc. The vertical distribution of the gas density is described by the equation of hydrostatic equilibrium. The solution of this equation is
\begin{equation}
	\rho(r,\,z)=\rho(r,\,0)e^{-\frac{z^2}{H^2}}.
\end{equation}
Angular velocity dependence on coordinates follows from the equation of centrifugal balance,
\begin{equation}
\Omega = \Omega_k\left(1 + \frac{z^2}{r^2}\right)^{-3/4}.\label{Eq:Omega_z}
\end{equation}
We calculate the vertical profiles of the gas temperature in the optically thick regions using approach of \citet{mb91} and taking into account heating by cosmic rays and radioactive elements according to \citet{dalessio98}. We assume that the temperature near the accretion disc's outer boundary is equal to the temperature of the interstellar medium, $15$~K, due to heating by external sources. 

Ionization fraction $x$ is calculated using the model of \citet{dudorov87}, as it was done in DK14. The equations include the thermal ionization of metals and hydrogen, shock ionization by cosmic rays, X-rays, radioactive elements, radiative recombinations and recombinations on the dust grains with mean size $a_d$. Evaporation of the dust grains is taken into account. It is considered that mean dust grain radius decreases with temperature due to the evaporation of volatiles at $T>150$~K and silicates at $T>1500$~K.

\subsection{Induction equation}
\label{Sec:Modif}

\subsubsection{Induction equation with the Hall effect and buoyancy}
\label{Sec:IndEq}
DK14 considered the induction equation with OD and MAD. Let us add two additional terms describing the buoyancy of the toroidal magnetic field ${\bf B}_t = \left(B_r, \, B_{\varphi}, \, 0\right)$ and the Hall effect

\begin{eqnarray}
	\frac{\partial {\bf B}}{\partial t} &=& \nabla\times\left(({\bf V} + {\bf V}_{\rm{mad}}) \times {\bf B}\right)  + \nabla\times\left(\eta_{\rm{o}}\left(\nabla\times{\bf B}\right)\right) +  \nonumber\\
	& &  \nabla\times\left({\bf V}_B \times {\bf B}_t\right) + \nabla\times\left(\frac{c}{4\pi en_e}\left(\left(\nabla \times{\bf B}\right) \times {\bf B}\right)\right),\label{Eq:Ind}
\end{eqnarray}
where stationary MAD velocity
\begin{equation}
	{\bf V}_{\rm{mad}} = \frac{\left(\left(\nabla\times{\bf B}\right) \times {\bf B}\right)}{4\pi R_{\rm{in}}}, \label{Eq:MADvelocity}
\end{equation}
$R_{\rm{in}} = \mu_{\rm{in}}n_{\rm{i}}n_{\rm{n}}\langle\sigma v\rangle_{\rm{in}}$~--~the friction coefficient for ion-neutral collisions, $\mu_{\rm{in}}$~--~the reduced mass for ion and neutral particles, $n_{\rm{i}}$~--~ ions number density, $n_{\rm{n}}$~--~neutral particles number density, $\langle\sigma v\rangle_{\rm{in}}= 2\times 10^{-9}\,\rm{cm}^3\,\rm{s}^{-1}$ \citep{spitzer_book}~--~the coefficient of momentum transfer in collisions between ions and neutrals.

The third term in (\ref{Eq:Ind}) is responsible for OD, 
\begin{equation}
	\eta_{\rm{o}} = \frac{c^2}{4\pi\sigma}\label{Eq:nu_m}
\end{equation}
~--~ohmic diffusivity, $c$~--~the speed of light, $\sigma_{\rm{e}}$~--~electrical conductivity,
\begin{equation}
	\sigma_{\rm{e}} = \frac{e^2n_{\rm{e}}}{m_{\rm{e}}\nu_{\rm{en}}},\label{Eq:sigma_e}
\end{equation}
where $e$~--~electron charge, $n_{\rm{e}}$~--~electrons number density, $m_{\rm{e}}$~--~the mass of the electron and $\nu_{\rm{en}}$~--~mean collision rate between electrons and neutral particles,
\begin{equation}
	\nu_{\rm{en}} = \langle\sigma v\rangle_{\rm{en}} n_{\rm{n}},
\end{equation}
where $\langle\sigma v\rangle_{\rm{en}}= 10^{-7}\,\rm{cm}^3\,\rm{s}^{-1}$ \citep{nakano84}~--~the rate of momentum transfer in electron-neutral collisions.

The fourth term at the right-hand side of (\ref{Eq:Ind}) describes the buoyancy of the toroidal magnetic field with the velocity ${\bf V}_B = \left(0,\, 0,\, V_B\right)$ \citep{dudorov89}. The fifth term describes the Hall effect (see, for example, \cite{balbus_book}).

In the approximation of \cite{ss73}, radial derivatives can be neglected comparing to vertical derivatives. In this case, equation (\ref{Eq:Ind}) has the following form in cylindrical coordinate system:
\begin{eqnarray}
  \frac{\partial B_r}{\partial t} &=& B_z \frac{\partial V_r}{\partial z} + \frac{\partial}{\partial z}\left[\left( \eta_{\rm{o}} + \frac{B_r^2+B_z^2}{4\pi R_{\rm{in}}}\right)\frac{\partial B_r}{\partial z}\right] +\nonumber\\
  & & \frac{\partial}{\partial z}\left(\frac{B_rB_{\varphi}}{4\pi R_{\rm{in}}}\frac{\partial B_{\varphi}}{\partial z}\right) + \frac{\partial}{\partial z}\left(\eta_h\frac{\partial B_{\varphi}}{\partial z}\right)\nonumber\\
  & & - \frac{\partial}{\partial z}\left(B_r V_B\right),\label{Eq:BrH}\\
  \frac{\partial B_{\varphi}}{\partial t} &=& B_{z}\frac{\partial V_{\varphi}}{\partial z} + B_r\frac{\partial V_{\varphi}}{\partial r} + \nonumber\\
  & & \frac{\partial}{\partial z}\left[\left( \eta_{\rm{o}} + \frac{B_{\varphi}^2+B_z^2}{4\pi R_{\rm{in}}}\right)\frac{\partial B_{\varphi}}{\partial z}\right] + \nonumber\\
  & & \frac{\partial}{\partial z}\left(\frac{B_rB_{\varphi}}{4\pi R_{\rm{in}}}\frac{\partial B_r}{\partial z}\right)  - \frac{\partial}{\partial z}\left(\eta_h\frac{\partial B_r}{\partial z}\right) \nonumber\\
  & & - \frac{\partial}{\partial z}\left(B_{\varphi}V_B\right),\label{Eq:BphiH}\\
  \frac{\partial B_z}{\partial t} &=& -\frac{1}{r}\frac{\partial}{\partial r}\left(rV_rB_z\right), \label{Eq:Bz}
\end{eqnarray}
where $V_{\varphi}=\Omega r$,
\begin{equation}
	\eta_{\rm{h}} = \frac{cB_z}{4\pi en_{\rm{e}}}\label{Eq:EtaHall}
\end{equation}
~--~the Hall coefficient.

Equation (\ref{Eq:BrH}) shows that $B_r$ is generated from $B_z$ due to accretion with the velocity $V_r(z)$. Differential rotation with the velocity $V_{\varphi}(r,\,z)$ generates $B_{\varphi}$ from $B_z$ and $B_r$, according to equation (\ref{Eq:BphiH}). The diffusion of the radial and azimuthal magnetic field components takes place in the vertical direction. Equation (\ref{Eq:Bz}) reflects that $B_z$ is frozen in gas.

Derivatives $\partial V_r/\partial z$ and $\partial B/\partial z$ can be replaced by finite differences $V_r / z$ and $B^{+} / z$, where $B^{+}$ is the value at height $z$. Equatorial symmetry requirement implies that $B_r(z)=-B_r(-z)$ and $B_{\varphi}(z)=-B_{\varphi}(-z)$. This means that $B_r(z=0)=0$ and $B_{\varphi}(z=0)=0$. In the steady-state case, equations (\ref{Eq:BrH}-\ref{Eq:BphiH}) are transformed to
\begin{eqnarray}
	-zV_rB_z - \eta_{\rm{h}}B_{\varphi}  &=& B_r\left(\eta + V_{B}z\right),\label{Eq:BrHSol}\\
	-\frac{1}{2}\Omega_kz^2\left( 3\frac{z}{r}B_z + B_r\right) + \eta_{\rm{h}}B_r &=& B_{\varphi}\left(\eta + V_{B}z\right),\label{Eq:BphiHSol}
\end{eqnarray}
where
\begin{equation}
\eta = \eta_{\rm{o}} + \eta_{\rm{a}}
\end{equation}
~--~sum of the ohmic diffusivity (\ref{Eq:nu_m}) and the ambipolar diffusivity
\begin{equation}
 \eta_{\rm{a}} = \frac{B^2}{4\pi R_{\rm{in}}},\label{Eq:eta_mad}
\end{equation} 
where $B^2 = B_r^2 + B_{\varphi}^2 + B_z^2$. The magnetic field components in the (\ref{Eq:BrHSol}-\ref{Eq:BphiHSol}) are evaluated at height $z$ above the mid-plane of the accretion disc. Upper index plus is omitted for simplicity.

The steady-state solution of equation (\ref{Eq:Bz}) is
\begin{equation}
B_z = B_{z0} \frac{\Sigma}{\Sigma_0},\label{Eq:BzFrozen}
\end{equation}
where $B_{z0}$ and $\Sigma_{0}$~--~the boundary values of the magnetic field and surface density. We assume that $B_{z0}(r_{\rm{out}}) = B_{\rm{ext}}(r_{\rm{out}})$ at the outer boundary of the accretion disc, where $B_{\rm{ext}}$ is the magnetic field of a protostellar cloud \citep{dudorov91}
\begin{equation}
	B_{\rm{ext}} = B_c \left(\frac{n_{\rm{ext}}}{n_c}\right)^{k_B},
\end{equation}
where $n_{\rm{ext}}$~--~gas number density at the accretion disc outer edge, $B_c$ and $n_c$ are the magnetic field and gas number density of the protostellar cloud from which a protostar with the accretion disc is formed. The typical values for the low-mass protostellar clouds $B_c=10^{-4}$~Gs, $n_c=10^4\,\rm{cm}^{-3}$. For the magnetostatic contraction of the protostellar disc $k_B=1/2$.

DK14 have shown that MAD may prevent the amplification of $B_z$ in the regions of low ionization fraction. In this case, $B_z$ can be estimated from the equality of $V_{\rm{mad}} = V_r$,
\begin{equation}
	B_z = \sqrt{4\pi R_{\rm{in}}r|V_r|}.\label{Eq:BzMAD}
\end{equation}

Equations (\ref{Eq:BrHSol}-\ref{Eq:BphiHSol}) are non-linear algebraic equations, since the coefficient of MAD $\eta_{\rm{a}}$ and the buoyancy velocity $V_B$ depend on the magnetic field. It should be noted that the coefficient $\eta_h$ can be either positive or negative depending on the sign of $B_z$, i.e. on the direction of the vertical magnetic field vector with respect to the angular velocity vector $\bf{\Omega}$. In our calculations, the rotation axis is directed along the $z$-axis, ${\bf \Omega}=\left(0,\, 0,\, \Omega>0\right)$. 

\subsubsection{The stationary velocity of the magnetic flux buoyancy}
\label{Sec:Vb}

DK14 used simple estimation of the azimuthal magnetic field intensity in the regions of effective generation. It was considered that the maximum value of $B_{\varphi}$ corresponds to the case when magnetic flux tubes (MFTs) rise with the Alfven speed. In present work, we investigate this problem more carefully.

Magnetic field splits into the MFTs due to the \citet{parker_book} instability in the regions where $B_{t}$ becomes larger than $B_z$. We assume that MFTs have the form of the torus with minor radius $a_{\rm{mft}}<<H$ and major radius $a_c\sim r$. Since the accretion discs of young stars are geometrically thin, $H \ll r$, then major radius $a_c\gg  a_{\rm{mft}}$. Thus, MFT can be treated locally as the thin cylinder. We do not take into account motion in $r$-direction and consider that MFTs rise in $z$-direction from the disc interior to its surface layers. There is pressure balance between the MFT and the external medium,
\begin{equation}
	P_{\rm{in}} + \frac{B^2_t}{8\pi} = P_{\rm{ext}},\label{Eq:Pbal}
\end{equation}
where $P_{\rm{in}}$~--~gas pressure inside the MFT, $P_{\rm{ext}}$~--~external gas pressure. In the case of effective heat exchange, when MFT temperature is equal to the surrounding gas temperature, it follows from (\ref{Eq:Pbal}) that the difference of MFT and surrounding medium densities is
\begin{equation}
	\Delta\rho=\rho_{\rm{ext}}-\rho_{\rm{in}} = \frac{B_t^2}{8\pi V_s^2},\label{Eq:DeltaRho}
\end{equation}
i.e. the gas inside the MFT has smaller density than the external one, $\rho_{\rm{in}} < \rho_{\rm{ext}}$. The buoyancy force 
\begin{equation}
F_{B}=-\Delta\rho g_z,\label{Eq:Fbuo1}
\end{equation}
causes the MFT to floating up ($g_z = -\Omega^2 z$ is the $z$-component of the stellar gravity).

In this section, we modify our accretion disc model by incorporating the buoyancy of the toroidal magnetic field in the induction equation. In order to calculate $B_r$ and $B_{\varphi}$, we need to specify the MFT rise velocity $V_B$. The stationary rise velocity of MFT is determined from the balance between the buoyancy and drag forces, $F_B$ and $F_D$. We obtain from  (\ref{Eq:DeltaRho}) and (\ref{Eq:Fbuo1})
\begin{equation}
	|F_B| = |g_z|\frac{B^2}{8\pi V_s^2}, \label{Eq:Fbuo2}
\end{equation}
where $B=B_t$~--~the magnetic field inside the MFT. The aerodynamic drag force is (see \cite{parker_book})
\begin{equation}
|F_{D}| = \frac{\rho_e V_B^2}{2}\frac{C_d}{\rho\pi a_{\rm{mft}}},\label{Eq:Fad}
\end{equation}
where $C_D$~--~drag coefficient, $\rho$~--~gas density inside the MFT. Typically, drag coefficient $C_D$ is of order of unity. In the turbulent medium, the drag force is \citep{pneuman72}
\begin{equation}
|F_t| = \frac{\pi\rho_e\left(\nu_t a_{\rm{mft}} V_B^3\right)^{1/2}}{\rho\pi a_{\rm{mft}}^2},\label{Eq:Ft}
\end{equation}
where $\nu_t$~--~turbulent viscosity. The forces in (\ref{Eq:Fbuo2}, \ref{Eq:Fad}, \ref{Eq:Ft}) are evaluated per unit mass of the MFT.

Equating the buoyancy force (\ref{Eq:Fbuo2}) and aerodynamic drag force (\ref{Eq:Fad}) with the help of equations for accretion disc scale height (\ref{Eq:H}) and angular velocity (\ref{Eq:Omega_k}), we obtain expression for the buoyancy velocity
\begin{equation}
	V_B = V_a\left(\frac{\pi}{C_D}\right)^{1/2}\left(\frac{a_{\rm{mft}}}{H}\right)^{1/2}\left(\frac{z}{H}\right)^{1/2},\label{Eq:V_b_a}
\end{equation}
where
\begin{equation}
	V_a = \frac{B}{\sqrt{4\pi\rho_e}}
\end{equation}
is the Alfven velocity. Expression (\ref{Eq:V_b_a}) shows that the typical MFT rise velocity for the tubes with $a_{\rm{mft}}\sim H$ is approximately equal to Alfven speed at height $z\sim H$.

Stationary MFT rise velocity in case of the turbulent drag follows from the equality of (\ref{Eq:Fbuo2}) and (\ref{Eq:Ft})
\begin{equation}
	V_B = V_a \left(\frac{\beta}{2}\right)^{-1/2}\alpha^{-1/3}\left(\frac{a_{\rm{mft}}}{H}\right)\left(\frac{z}{H}\right)^{2/3},\label{Eq:Vb_t}
\end{equation}
where $\beta = 8\pi\rho V_s^2/B^2$ is the plasma parameter, and turbulent viscosity is evaluated according to \citet{ss73}
\begin{equation}
	\nu_t = \alpha V_s H.\label{Eq:nu_t}
\end{equation}
In order to take into account the buoyancy effect, we substitute (\ref{Eq:V_b_a}) or (\ref{Eq:Vb_t}) into equations (\ref{Eq:BrHSol}-\ref{Eq:BphiHSol}).

\section{Fossil magnetic field intensity and geometry in accretion discs}
\label{Sec:Results}
In this section, we investigate the fossil magnetic field of accretion discs. We use the following fiducial values of parameters: stellar mass $M=1\,\rm{M}_{\odot}$, accretion rate $10^{-7}\,\rm{M}_{\odot}\,\rm{yr}^{-1}$, turbulence parameter $\alpha=0.01$, stellar luminosity $L_{\rm{s}} = 1\,\rm{L}_{\odot}$, stellar radius $R_{\rm{s}}=2\,\rm{R}_{\odot}$, mean magnetic field strength at the stellar surface $B_{\rm{s}}=2$~kGs, cosmic rays ionization rate $\xi_0=10^{-17}\,\rm{s}^{-1}$ and attenuation length $R_{\rm{CR}}=100\,\rm{g}\,\rm{cm}^{-2}$, X-ray luminosity $L_{\rm{XR}}=10^{30}\,\rm{erg}\,\rm{s}^{-1}$, mean dust particle radius $a_{\rm{d}}=0.1\,\mu\rm{m}$. The radial profiles of the magnetic field components $B_r$ and $B_{\varphi}$ are evaluated at height $z=0.5\,H$. We depict the absolute values of the magnetic field components in all figures.

The system of non-linear algebraic model equations  (\ref{Eq:AngMom}-\ref{Eq:Omega_z}, \ref{Eq:BrHSol}-\ref{Eq:BzFrozen}, \ref{Eq:BzMAD}) is solved numerically in three steps.  At the first step, equations (\ref{Eq:AngMom}-\ref{Eq:H}) are solved. At the second step, the ionization subsystem (see equations (41, 47, 48) in DK14) is solved using the iterative Newton's method with the relative accuracy $10^{-4}$. Finally, the magnetic field components are determined by solving equations (\ref{Eq:BrHSol}-\ref{Eq:BphiHSol}, \ref{Eq:BzMAD}-\ref{Eq:BzFrozen}) with the help of the Newton's method.

\subsection{Accretion disc structure}

\begin{figure}
\center{\includegraphics[width=\columnwidth]{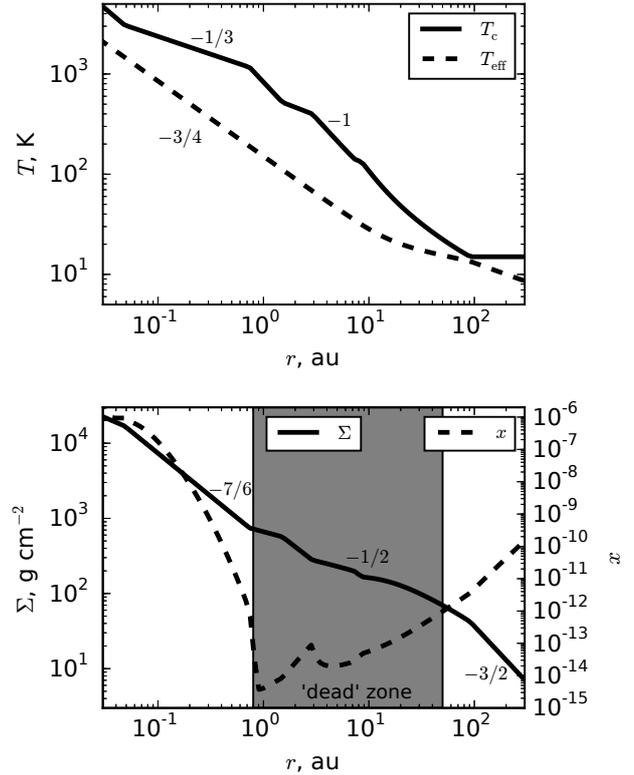}}
\caption{{\it Top}: The effective (dashed line) and mid-plane (solid line) temperatures versus distance $r$. {\it Bottom}: gas surface density (black line, left $y$-axis) and ionization fraction (gray line, right $y$-axis) in the mid-plane of the disc versus distance $r$. Numbers show characteristic slopes.}
\label{Fig:1}
\end{figure}

Before starting to analyse the magnetic field, we briefly discuss the structure of the disc. In Fig. \ref{Fig:1}, we show the radial profiles of effective temperature $T_{\rm{eff}}$ and mid-plane gas temperature $T_c$ (top panel), gas surface density $\Sigma$ and mid-plane ionization fraction $x$ (bottom panel).

Fig. \ref{Fig:1} shows that the disc effective temperature decreases with the distance $r$ as $T_{\rm{eff}}\propto r^{-3/4}$. Mid-plane temperature falls down with the distance more rapidly, $T\sim r^{-1}$ at $r>1$~au.  It is equal to $5200$~K near the accretion disc inner edge, $r_{\rm{in}}\sim 0.03$~au. The optical thickness of the disc is less than unity, and the temperature is equal to the temperature of molecular cloud cores $15$~K at $r>100$~au. The gas surface density decreases with the distance from the value $\sim 2\times 10^4\,\rm{g}\,\rm{cm}^{-2}$ at the inner edge of the disc $r_{\rm{in}}$ to $\sim 10\,\rm{g}\,\rm{cm}^{-2}$ at the disc's outer edge, $r_{\rm{out}}\sim 300$~au. 

The ionization fraction profile is non-monotonic. In region $r<0.8$~au with $T>10^3$~K ionization fraction is determined by the thermal ionization. In the outer part of the disc, the ionization fraction increases with the distance, as cosmic rays ionization rate grows as the density falls down. The ionization fraction reaches extremely low value $\sim 10^{-14}$ at $r\sim 1$~au. In this region, thermal ionization does not operate, and cosmic rays poorly ionize gas because of high density, $\Sigma > R_{\rm{CR}}$. The small peak in the $x(r)$ profile at $r\sim 3$~au corresponds to the volatile grains evaporation. The region of low ionization fraction and efficient magnetic diffusion (`dead' zone) is located between $r\simeq 0.8$~au and $r\simeq 50$~au (gray region in Fig. \ref{Fig:1}). The ionization fraction is less than $\sim 10^{-12}$ inside the `dead' zone.

\subsection{The role of MHD effects}

To determine the conditions under which Ohmic diffusion, magnetic ambipolar diffusion, and the Hall effect are important, we compare magnetic diffusivities, $\eta_{\rm{o}}$ and $\eta_{\rm{a}}$, with the Hall coefficient $\eta_h$. Dividing (\ref{Eq:EtaHall}) by (\ref{Eq:nu_m}), we find that
\begin{equation}
	\frac{\eta_{\rm{h}}}{\eta_{\rm{o}}} = \omega_{\rm{e}}\tau_{\rm{en}},\label{Eq:eta_h_nu_m}
\end{equation}
where
\begin{equation}
	\omega_{\rm{e}} = \frac{eB_z}{m_{\rm{e}}c}
\end{equation}
~--~the cyclotron frequency for electrons, $\tau_{\rm{en}} = \nu_{\rm{en}}^{-1}$~--~the mean time of electron-neutral collisions. The relation (\ref{Eq:eta_h_nu_m}) shows that the Hall effect is more efficient than Ohmic diffusion in the magnetized plasma, $\omega_{\rm{e}}\tau_{\rm{en}}>1$. For typical values $B_z=0.1$~Gs and $n = 10^{14}\,\rm{cm}^{-3}$ at $r=1$~au
\begin{equation}
	\omega_{\rm{e}}\tau_{\rm{en}} = 0.18\left(\frac{B_z}{0.1\,\rm{Gs}}\right)\left(\frac{n}{10^{14}\,\rm{cm}^{-3}}\right)^{-1},\label{Eq:we_tau}
\end{equation}
i.e. plasma is not magnetized, and the Hall effect is not efficient at $r=1$~au. 

Division of (\ref{Eq:eta_mad}) by (\ref{Eq:nu_m}) yields
\begin{equation}
	\frac{\eta_{\rm{a}}}{\eta_{\rm{o}}} = \left(\omega_{\rm{e}}\tau_{\rm{en}}\right)\left(\omega_{\rm{i}}\tau_{\rm{in}}\right)\frac{m_{\rm{i}}+m_{\rm{n}}}{m_{\rm{n}}},\label{Eq:eta_mad_nu_m}
\end{equation}
where
\begin{equation}
	\omega_{\rm{i}} = \frac{eB_z}{m_{\rm{i}}c}
\end{equation}
~--~the cyclotron frequency for ions, $\tau_{\rm{in}} = \nu_{\rm{in}}^{-1}$~--~the mean time of ion-neutral collisions, $m_{\rm{n}}$~--~the mass of the neutral particle. For the typical scales
\begin{equation}
	\frac{\eta_{\rm{a}}}{\eta_{\rm{o}}} = 5.6\times 10^{-4}\left(\frac{B_z}{0.1\,\rm{Gs}}\right)\left(\frac{n}{10^{14}\,\rm{cm}^{-3}}\right)^{-1}.\label{Eq:wewi_tau}
\end{equation}

\begin{figure}
\center{\includegraphics[width=\columnwidth]{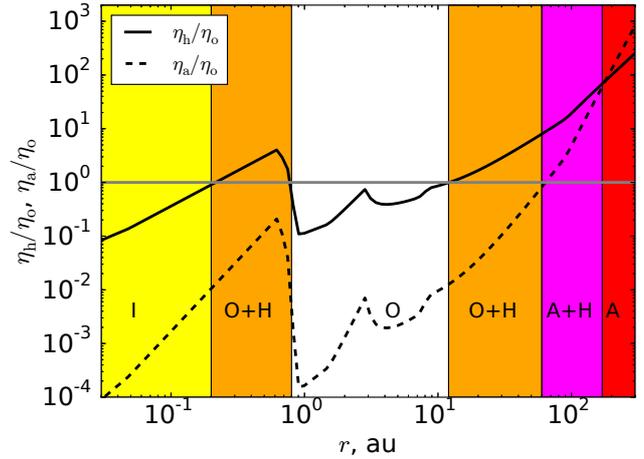}}
\caption{The radial profiles of the relations $\eta_{\rm{h}} / \eta_{\rm{o}}$ and $\eta_{\rm{a}} / \eta_{\rm{o}}$. The colour fillings show the regions where considered MHD effects are important (I~--~frozen-in field, O~--~Ohmic diffusion, A~--~magnetic ambipolar diffusion, H~--~Hall effect). (The color version of this figure is available in the online journal).}
\label{Fig:2}
\end{figure}

In Fig. \ref{Fig:2}, we plot the radial profiles of the relations (\ref{Eq:we_tau}) and (\ref{Eq:wewi_tau}) calculated with the fiducial parameters. Fig. \ref{Fig:2} shows that the electrons are magnetized, $\omega_{\rm{e}}\tau_{\rm{en}}>1$, in the outer part of the disc, $r>20$~au. The Hall effect develops together with Ohmic diffusion at $20\,\rm{au}<r<60$~au (the orange region in Fig.\ref{Fig:2}), and together with magnetic ambipolar diffusion in region $60\,\rm{au}<r<150$~au (the magenta region in Fig. \ref{Fig:2}). The MAD is the dominant effect at $r>160$~au (red region in Fig. \ref{Fig:2}). The Hall effect and Ohmic diffusion are also important in the small region $0.2\,\rm{au}<r<0.8$~au of refractory grains evaporation. In the innermost part of the disc, $r<0.2$~au, the magnetic field is frozen-in.

\subsection{The influence of Ohmic and magnetic ambipolar diffusion}
\label{Sec:ResBu}
\begin{figure}
\center{\includegraphics[width=\columnwidth]{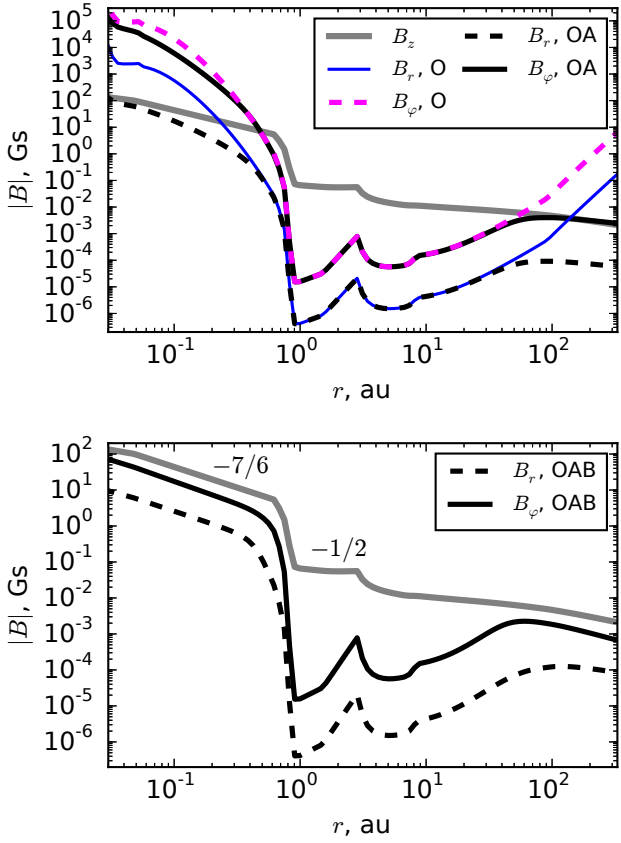}}
\caption{{\it Top}: the radial profiles of $B_z$ (gray line), $B_r$ and $B_{\varphi}$ calculated taking into account OD (blue line and magenta dashed line, respectively), and that, calculated taking into account OD and MAD (black dashed and solid lines, respectively). {\it Bottom}: the radial profiles of $B_z$ (gray line), $B_r$ and $B_{\varphi}$ calculated taking into account  OD, MAD and buoyancy. Numbers show characteristic slopes. (The color version of this figure is available in the online journal).}
\label{Fig:3}
\end{figure}

In Fig. \ref{Fig:3}, we show the radial profiles of the $B_z$, $B_r$ and $B_{\varphi}$ calculated with the fiducial parameters. In the top panel of Fig. \ref{Fig:3} two cases are shown: with OD only and with both OD and MAD.

According to Fig. \ref{Fig:3}, $B_z$ decreases with the distance $r$. At the inner edge of the disc, $r_{\rm{in}}$, $B_z\sim 345$~Gs . The magnetic field is frozen-in and $B_z\propto \Sigma \propto r^{-7/6}$ at $r<0.8$~au. At $r\simeq 0.8$~au, $B_z$ drops by two orders of magnitude, from $\sim 10$~Gs to $\sim 0.1$~Gs. Magnetic ambipolar diffusion reduces $B_z$ comparing to the frozen-in intensity in the region of low ionization fraction, $r>0.8$~au. The radial profile of $B_z$ is determined by dependence (\ref{Eq:BzMAD}) in this region, $B_z \propto r^{-1/2}$. Small peak in $B_z(r)$ at $r\sim 3$~au coincides with the ionization fraction peak caused by volatile grains evaporation.

According to the top panel of Fig. \ref{Fig:3}, the intensity of the toroidal magnetic field is extremely large near the inner edge of the disc, because OD and MAD are not efficient in this region. The extreme growth of  $B_r$ and $B_{\varphi}$ also takes place near the outer edge of the disc, if MAD is not taken into account. Magnetic ambipolar diffusion limits the generation of $B_r$ and $B_{\varphi}$ in outer part of the disc, so that the azimuthal magnetic field component intensity is comparable with $B_z$, while $B_r$ is ten times lower. 

The radial and azimuthal magnetic field components are small comparing to the $B_z$ due to OD and MAD inside the `dead' zone, $0.8\,\rm{au}<r<50$~au.

\subsection{The influence of buoyancy}
\label{Sec:ResBu}

In the bottom panel of Fig. \ref{Fig:3} we show the radial profiles of $B_z$, $B_r$ and $B_{\varphi}$ calculated under the influence of OD, MAD, and buoyancy. The MFT radius is equal to $1\,H$, the turbulent drag is taken into account.

According to the bottom panel of Fig. \ref{Fig:3}, $B_{\varphi}$ is nearly two times lower than $B_z$ in the region of frozen-in magnetic field, $r<0.8$~au. The intensity of the radial magnetic field component is by order of magnitude lower than the $B_z$. The radial profiles of $B_r$ and $B_{\varphi}$ are similar to $B_z(r)$ in this region. The buoyancy also operates in the outer region of the disc, so that $B_{\varphi}$ is nearly two times lower than $B_z$ at $r>50$~au.

We calculated the toroidal magnetic field of the disc for different MFT radii, $a_{\rm{mft}}=(0.01-1)$~H. Calculations show that the more the MFT radius, the less the toroidal magnetic field intensity in the regions of the effective generation. The buoyancy velocity (\ref{Eq:Vb_t}) grows with the MFT radius. Ten times growth of the MFT radius leads to nearly two times decrease of the $B_r$ and $B_{\varphi}$, both in the inner and in the outer regions of the disc. 

\subsection{Hall effect influence on the fossil magnetic field}
\label{Sec:ResHall}
\begin{figure}
\center{\includegraphics[width=\columnwidth]{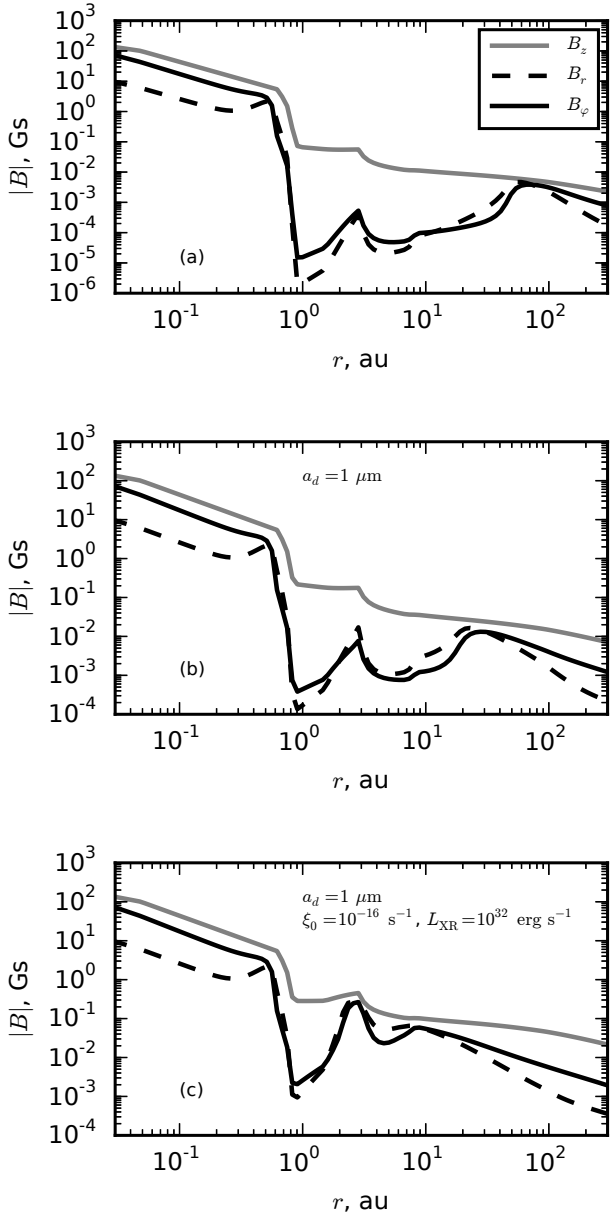}}
\caption{The radial profiles of the magnetic field components calculated taking into account Ohmic diffusion, magnetic ambipolar diffusion, buoyancy and the Hall effect. {\it Panel a}: fiducial parameters, {\it panel b}: $a_d=1\,\mu\rm{m}$, {\it panel c}: $a_d=1\,\mu\rm{m}$, $\xi_0=10^{-16}\,\rm{s}^{-1}$, $L_{\rm{XR}}=10^{32}\,\rm{erg}\,\rm{s}^{-1}$.}
\label{Fig:4}
\end{figure}

In this section, we investigate how the Hall effect changes the magnetic field in the disc. In Fig. \ref{Fig:4}, we show the radial profiles of the magnetic field components calculated taking into account OD, MAD, buoyancy and the Hall effect for the case $B_z>0$. The panel (a) corresponds to the calculation with the fiducial parameters. In the panels (b) and (c), we plot the dependences with different dust grain and ionization parameters.

Fig. \ref{Fig:4}a shows that the Hall effect changes relation between $B_r$ and $B_{\varphi}$, i.e. it changes the geometry of the magnetic field in the regions where $\omega_{\rm{e}}\tau_{\rm{en}}>1$ (orange regions in Fig. \ref{Fig:2}). The intensity of the radial magnetic field component is comparable with $B_z$ and $B_{\varphi}$ due to the Hall effect in the regions where electrons are magnetized. We did not investigate in details case $B_z<0$. The behaviour of the magnetic field in the presence of the Hall effect is expected to be oscillatory \citep{shalybkov97}. We believe that full 3D global simulations of the accretion discs will confirm this conclusion.

Fig. \ref{Fig:4}b and \ref{Fig:4}c show that the variation of the ionization parameters leads to the change of the `dead' zone size, while the magnetic field geometry is not changed. For example, ten times increase of the dust grain radius (Fig. \ref{Fig:4}b) leads to ten times growth of the ionization fraction. In this case, the extent of the `dead' zone two times shrinks comparing to the fiducial case (it is located at $0.8$~au$<r<23$~au). The `dead' zone is situated between 0.8~au and 8~au in the case of the enhanced rates of the ionization by cosmic rays and X-rays (Fig. \ref{Fig:4}c).

\begin{figure}
\center{\includegraphics[width=\columnwidth]{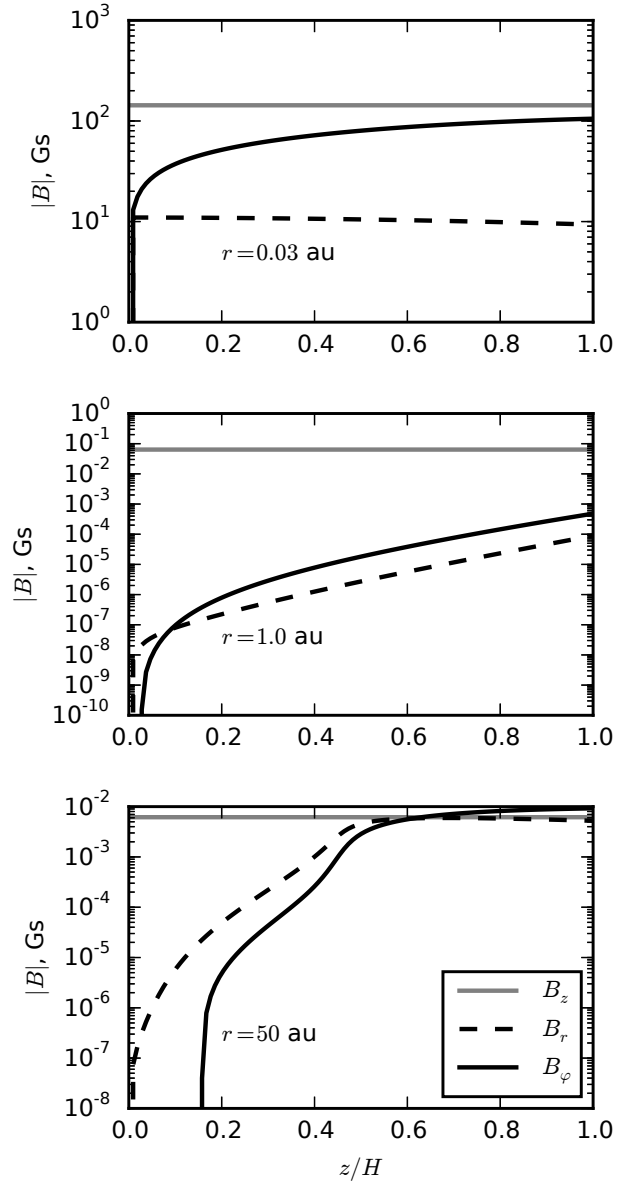}}
\caption{The vertical profiles of the magnetic field components calculated taking into account Ohmic diffusion, magnetic ambipolar diffusion, buoyancy and the Hall effect at $r=0.03$~au (top panel), $r=1$~au (middle panel), and $r=50$~au (bottom panel).}
\label{Fig:5}
\end{figure}

In Fig. \ref{Fig:5}, we plot the vertical profiles of the fossil magnetic field components at three radial distances from the star, 0.03~au, 1~au, and 50~au. The vertical magnetic field component does not depend on $z$-coordinate. The radial and azimuthal magnetic field components increase with the height above the mid-plane, where $B_r=0$ and $B_{\varphi}=0$. Fig. \ref{Fig:5} confirms previous results. The magnetic field is quasi-azimuthal at $z>0.5\,H$ and $r=0.03$~au, quasi-vertical inside the `dead' zones, $r=1$~au, and quasi-radial at $r=50$~au,  $z>0.5\,H$.

It should be noted that the Hall effect is non-dissipative, while the OD and MAD are the diffusion ones. Fig. \ref{Fig:2} shows that the Hall effect operates {\it together} with OD and/or MAD in the regions where the plasma is magnetized. According to Fig. \ref{Fig:2} and \ref{Fig:4}, the Hall effect is the most important near the `dead' zone boundaries. The Hall effect leads to the transformation of the magnetic field geometry, while OD and/or MAD constraint the generation of the magnetic field and lead to magnetic flux decrease in these regions. The magnetic field gets quasi-radial geometry due to the Hall effect.

\subsection{Fossil magnetic field in the accretion discs with different accretion rates}
\label{Sec:Mdot}
\begin{figure*}
\center{\includegraphics[width=0.85\textwidth]{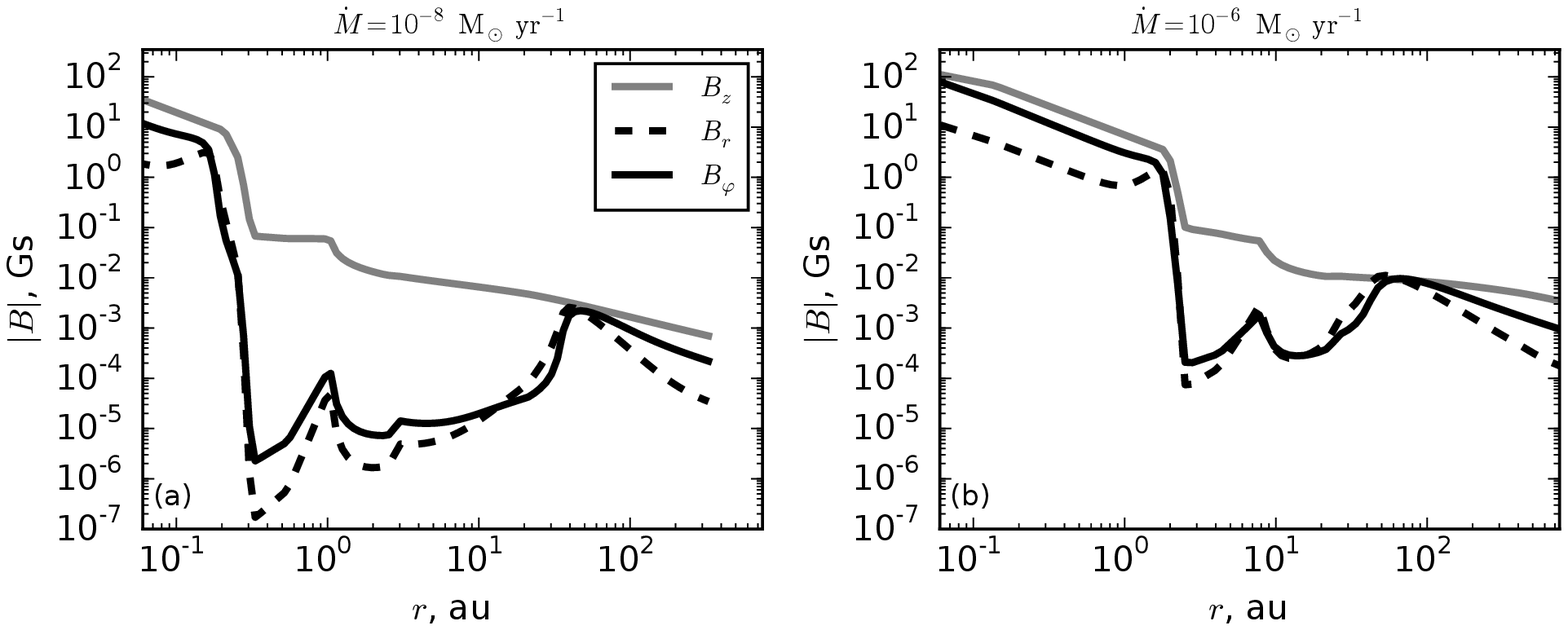}}
\caption{The radial profiles of the magnetic field components calculated taking into account OD, MAD, buoyancy, and the Hall effect. Gray line: $B_z$, black solid line: $B_{\varphi}$, black dashed line: $B_r$. {\it Panel a}: $\dot{M}=10^{-8}\,\rm{M}_{\odot}\,\rm{yr}^{-1}$; {\it panel b}: $\dot{M}=10^{-6}\,\rm{M}_{\odot}\,\rm{yr}^{-1}$.}
\label{Fig:6}
\end{figure*}

In this subsection, we analyse how the geometry and the intensity of the fossil magnetic field depend on the accretion rate. Fig. \ref{Fig:6} shows the radial profiles of the magnetic field components calculated with the accretion rates $10^{-8}\,\rm{M}_{\odot}\,\rm{yr}^{-1}$ (panel a) and $10^{-6}\,\rm{M}_{\odot}\,\rm{yr}^{-1}$ (panel b), taking into account OD, MAD, buoyancy, and the Hall effect. 

Fig. \ref{Fig:6} shows that the fossil magnetic field geometry does not depend on the accretion rate. The magnetic field is quasi-vertical, $B_z\gg (B_r,\,B_{\varphi})$, inside the `dead' zones, it is quasi-azimuthal, $B_{\varphi}\sim B_z$, near the inner edge of the disc, quasi-radial, $B_r\sim B_z$, outside the `dead' zone at $50-100$~au, and quasi-azimuthal at farther distances.  The `dead' zone is more extended in the disc with the higher accretion rate. This is due to the more the accretion rate, the more massive the accretion disc. As the accretion rate decreases during the disc evolution, the `dead' zone shrinks, and it is shifted to the star. For example, the inner edge of the `dead' zone is located at $r\sim 2$~au in case $\dot{M}=10^{-6}\,\rm{M}_{\odot}\,\rm{yr}^{-1}$, and at $r\sim 0.2$ in case $\dot{M}=10^{-8}\,\rm{M}_{\odot}\,\rm{yr}^{-1}$. At any given distance from the star, the magnetic field intensity is lower in the disc with the smaller accretion rate. For example, $B_z(3\,\rm{au})\sim 0.1$~Gs and $B_{\varphi}(3\,\rm{au})\sim 2\times 10^{-4}$~Gs in the disc with $\dot{M}=10^{-6}\,\rm{M}_{\odot}\,\rm{yr}^{-1}$, and $B_z(3\,\rm{au})\sim 0.01$~Gs, $B_{\varphi}(3\,\rm{au})\sim 2\times 10^{-5}$~Gs in the disc with $\dot{M}=10^{-8}\,\rm{M}_{\odot}\,\rm{yr}^{-1}$.

\begin{table*}
\small
\caption{The fossil magnetic field in the accretion disc at typical distances with different accretion rates.}
\centering
\begin{tabular}{lccc}
\hline 
$\dot{M}$, $10^{-7}\,\rm{M}_{\odot}\,\rm{yr}^{-1}$ & $B_z(r_{\rm{in}})$, Gs & $B_z(3\,\mbox{au})$, Gs & $B_z(r_{\rm{out}})$, Gs \\ 
(1) & (2) & (3) & (4) \\ 
\hline 
$10$ & $320\,(0.01\,\rm{au})$ $(\beta=2500)$ & $0.09$ $(\beta=1.4\times 10^4)$ & $0.0033\,(710\,\rm{au})$ $(\beta=6)$ \\ 
$1$ &  $140\,(0.03\,\rm{au})$ $(\beta=360)$ & $0.03$ $\beta=1.9\times 10^4)$ & $0.0020\,(340\,\rm{au})$ $(\beta=20)$ \\ 
$0.1$ & $40\,(0.05\,\rm{au})$ $(\beta=90)$ & $0.01$ $(\beta=3\times 10^4)$ & $0.0012\,(150\,\rm{au})$ $(\beta=50)$ \\ 
\hline 
\end{tabular} 
\label{Tab:1}
\end{table*}

In Table \ref{Tab:1}, we show the values of $B_z$ and related $\beta$ values at typical distances in the disc with different accretion rates. Table \ref{Tab:1} shows that the magnetic field in the `dead' zones and in the regions of frozen-in field decreases when the accretion rate decreases. If the accretion rate ten times drops, then magnetic field in the disc nearly 2-2.5 times reduces. The magnetic field strength lies in the range $\sim(0.01-0.1)$~Gs at $3$~au. These values are in agreement with the measurements of the remnant magnetizations of meteorites in the Solar system \citep{levy78, fu14}.The plasma parameter at given distance increases with the decreasing mass accretion rate. It also varies with the distance. For example, $\beta=360$ at the inner edge of the disc, $\beta=1.9\times 10^4$ at 3 au, and $\beta=20$ at the outer edge of the disc with the fiducial parameters. Therefore, the approximation of the initial constant plasma beta for the whole disc is very rough.

\section{Synthetic polarization maps}
\label{sec:synth}

In this section, we make the synthetic maps of the polarized continuum dust emission using the accretion disc structure with the fiducial parameters. It is considered that the polarization comes from the dust grains alignment with the magnetic field direction in the disc. For the calculations, we use the monochromatic dust opacities of chemically inhomogeneous dust aggregates with `normal' silicate mineralogy calculated by \citet{Semenov_etal2003} for $T_{\rm{d}}<120$~K. We present results for the case when OD, MAD, buoyancy, and the Hall effect are taken into account (see Fig.~\ref{Fig:4}a). 

\subsection{Dust temperature calculations}
\label{sec:hyperion}

The dust temperature was calculated with \textsc{hyperion} Monte~Carlo radiative transfer code version 0.9.7 \citep{Robitaille2011}. The calculations were performed using spherical polar grid containing 110, 550, and 2 cells in the radial, polar and azimuthal direction, respectively.

For $T_{\rm{d}}$ calculations, we took the gas density distribution from the accretion disc model. The dust density distribution was obtained assuming that the dust and gas are well mixed, and gas-to-dust mass ratio is 100. The values of gas density in the points of \textsc{hyperion} spherical polar grid were linearly interpolated from the cylindrical grid used in the disc model.

To take into account dust sublimation, we have set the dust density to zero at the cylindrical radii lower than the dust sublimation radius $R_{\rm{sub}}=0.075$~au. The value of $R_{\rm{sub}}$ was obtained with the analytic expression given by \citet{Whitney_etal2004}:

\begin{equation}
    R_{\rm{sub}}=R_{\rm{s}} \left( \frac{ T_{\rm{sub}} }{T_{*}} \right) ^{-2.085},
	\label{eq:quadratic}
\end{equation}

\noindent where $T_{*}$~--~stellar effective temperature of 4085~K, $T_{\rm{sub}}$~--~dust sublimation temperature 1500 K.

\textsc{hyperion} computes the dust equilibrium temperature with \citet{Lucy1999} iterative method. The convergence criteria was similar to the one given in Section 4.2.2 of \citet{Robitaille2011}. We set the number of photons packets of $10^7$ per iteration (an average of 80 packets per grid cell). We used the partial diffusion approximation and modified random walk procedure \citep{Min_etal2009,Robitaille_etal2010} to speed up the calculations in the optically thick disc regions.

\begin{figure}
	\includegraphics[width=\columnwidth]{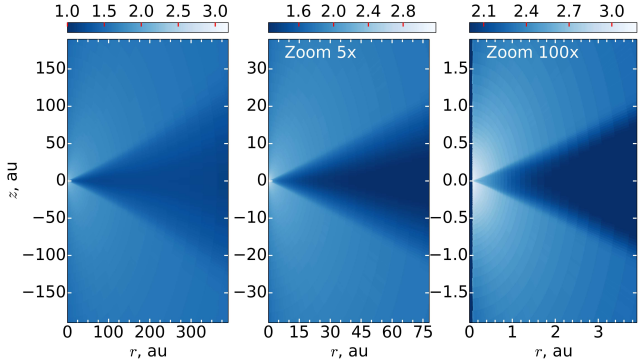}
    \caption{The distribution of the dust temperature for the fiducial disc model. Each image is a vertical slice through the half of the disc. The colour bars denote $\log_{10}$ of the dust temperature in K. (The color version of this figure is available in the online journal).}
    \label{Fig:7}
\end{figure}

The distribution of $T_{\rm{d}}$ obtained for the fiducial disc model is shown in Fig.~\ref{Fig:7}. It is seen that the opaque inner disc regions cast shadow on the equatorial disc plane. Thus, for example, $T_{\rm{d}}$ at $r=3$~au increases from $\sim50$~K at $z=0$~au to $220$~K at $|z|\sim1$~au (see right panel of Fig.~\ref{Fig:7}). The difference in $T_{\rm{d}}$ between disc regions at different $|z|$ becomes smaller with increasing $r$. $T_{\rm{d}}$ at $r=200$~au varies from 30 to 45~K with varying $z$. On average, $T_{\rm{d}}$ obtained for the disc models with $\dot{M}=10^{-8}$ and $\dot{M}=10^{-6}$~M$_{\sun}$~yr$^{-1}$ is higher by $\sim5$ and lower by $\sim10$ per cent, respectively, comparing with the fiducial model.

\subsection{Continuum polarization}
\label{sec:lime}

To obtain the synthetic maps of polarized continuum emission, we used the \textsc{lime} (Line Modelling Engine) code \citep{Brinch2010}. \textsc{lime} is the Monte Carlo code that allows to perform the molecular excitation and line radiative transfer calculations in far-infrared and (sub)mm wavelength regions. \citet{Padovani_etal2012} extended \textsc{lime} with the procedure to calculate the Stokes parameters \citep[see e.g.][]{Frau_etal2011}\footnote{Expressions for the Stokes parameters given by \citet{Padovani_etal2012} and in references therein should be corrected according to \citet{Planck_Collaboration2015}.}. With this procedure the polarization is assumed to be due to the grains alignment with the magnetic field. The stokes parameters are computed integrating the continuum source function along the line of sight taking into account the dust temperature variations and the variations of the magnetic field vector. The grain alignment efficiency and absorption cross-section are described by the single parameter $\alpha_{\rm{max}}$ that also determines the upper limit on the polarization degree, $p$, of continuum emission. We set $\alpha_{\rm{max}}=0.01$ which gives upper limit $p=1\%$ that is comparable with the limits obtained from the observations of the T~Tauri star discs \citep{Hughes_etal2013}. We assume that the dust (sub)mm emission scattering is negligible. The scattering at (sub)mm wavelengths is effective only in the presence of large 100~$\mu$m sized grains \citep[see e.g.][]{Kataoka2015} which are not considered in our disc models. Thus, we do not take into account the effect of disc inclination considered by \citet{Yang_etal2016}.

\textsc{lime} uses 3D spatial grid for the integration of the radiation transfer equation along a number of discrete rays. The positions of grid points are randomly sampled from the uniform or weighted distribution. We used the \textsc{lime} grid of $1\,200\,000$ points which were distributed mainly in the disc regions with high density and high gradients of magnetic field. Among these points, $500\,000$ points were distributed logarithmically and uniformly on $r$ and $z$, respectively. The positions of other $100\,000$ points were randomly selected from the distribution with the weights of the form $\left( \rho_{\rm{d}}/\rho_{\rm{d\,max}} \right)^{0.3}$, where $\rho_{\rm{d}}$~--~dust density at a given point and $\rho_{\rm{d\,max}}$~--~maximum dust density \citep[similar weights were used by e.g.][]{Douglas_etal2013}. The values of $\rho_{\rm{d\,max}}$ are $10^{-9.6}$, $10^{-8.8}$ and $10^{-8.3}$ g~cm$^{-3}$ for the disc models with $\dot{M}=10^{-8}$, $\dot{M}=10^{-7}$ and $\dot{M}=10^{-6}$~M$_{\sun}$~yr$^{-1}$, respectively. Six sets of $100\,000$ points were uniformly distributed within the intersected disc regions with $|z|<0.0001H$, $|z|<0.001H$, $|z|<0.01H$, $|z|<0.1H$, $|z|<H$ and $|z|<5H$, respectively.

As input physical parameters for the continuum polarization calculations we used the strength of the magnetic field from the disc model and the dust temperature and density obtained in Section~\ref{sec:hyperion}. The values of ${\bf B}$ and $T_{\rm{d}}$ in the \textsc{lime} grid points were obtained by the linear interpolation of the values in spatial grids of the disc and \textsc{hyperion} models.

\begin{figure}
	\includegraphics[width=\columnwidth]{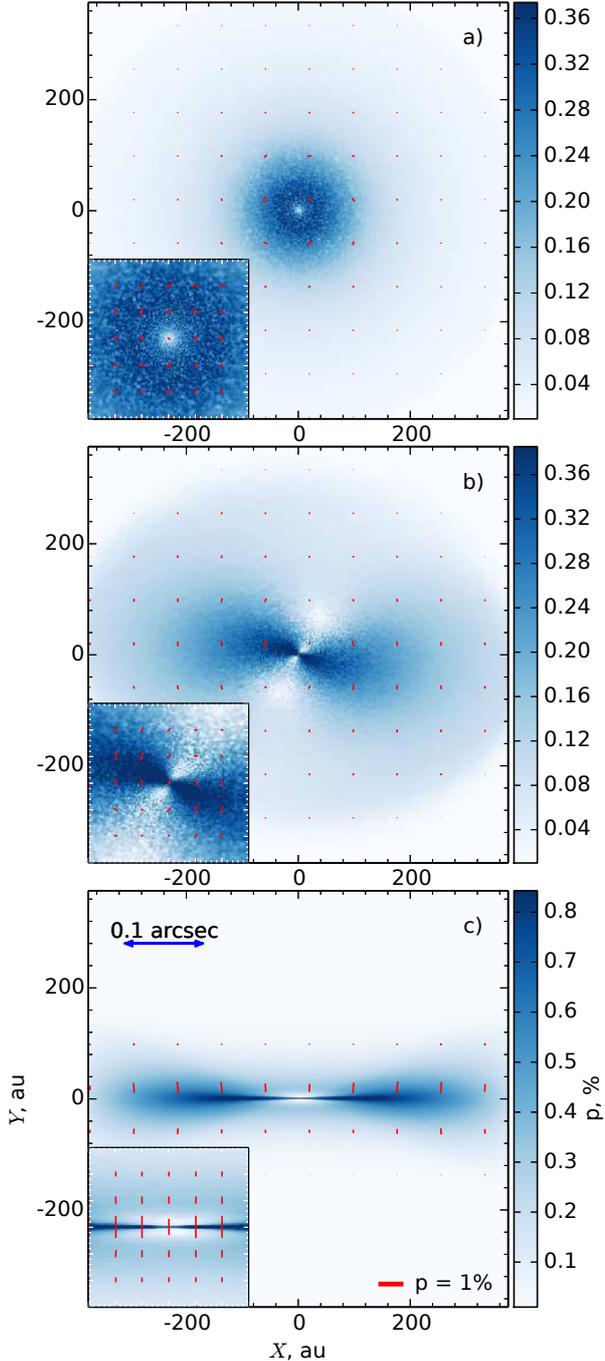}
    \caption{Polarization maps calculated for the disc with fiducial parameters. From top to bottom: disc inclination is 0, 45, $90\degr$. Sub-panels show zoomed $120\times120$~au central map region. The distance between major axes ticks on sub-panels is of 20~au. The colour bars denote polarization degree, $p$, and red lines denote polarization vectors flipped by $90\degr$. The red scale bar is consistent with $p=1\%$, the length of blue scale bar gives the angular scale of 1 arcsec at the distance of 140 pc. (The color version of this figure is available in the online journal).}
    \label{Fig:8}
\end{figure}

On the polarization map calculated at 237 GHz (1.3 mm) for the face-on disc with fiducial parameters, it is seen that at $r>15$~au $p$ increases up to $\sim0.35$ per cent with decreasing distance from the disc centre (see Fig.~\ref{Fig:8}a) due to increasing $B_r$ and $B_{\phi}$. The low values of $p$ and the orientation of the polarization vectors are consistent with almost vertical magnetic field. There is an apparent disc region within $r\lesssim15$~au, a polarization `hole', where $p$ is lower by factor of 2 comparing with neighbour disc regions with $r>15$~au. The appearance of this `hole' is related with the `dead' zone where the magnetic field is predominantly vertical and, thus, the projection of the magnetic field vector on the sky plane ($X$,$Y$) is minimum. The `hole' disappears for the disc inclination $i\gtrsim20\degr$.

The synthetic polarization map for the disc inclined on $45\degr$ reveals the asymmetric structure of $p$ distribution (see Fig.~\ref{Fig:8}b). Within the map region with $|X|<200$ and $|Y|<200$~au, one can distinguish two lobes that are approximately along $Y$ axis and two lobes that are approximately along $X$ axis. On average, $p$ values within lobes elongated along $Y$ are lower by a factor of $\sim3$ comparing with $p$ within the lobes elongated along $X$ axis. The asymmetry of $p$ distribution is due to the orientation of $\bf{B}$ towards the disc rotation direction. The asymmetry vanishes out for $i\lesssim30\degr$ and $i\gtrsim85\degr$.

In the case of disc seen edge-on, there is a narrow region with $|Y|<5$ and $|X|<100$~au on the polarization map where $p$ is maximum (see Fig.~\ref{Fig:8}c). This region has sharp borders where $p$ rapidly decreases from $\sim1$ to $\sim0.3\%$ with increasing $|Y|$. The presence of this narrow region with high $p$ is related with the variations of the magnetic field strength close to the disc mid-plane. $B_r$ and $B_{\phi}$ are much smaller than $B_z$ at $z\lesssim0.2H$ and rapidly increase at $z>0.2H$ (see Fig.\ref{Fig:5}) that leads to the variations of $\bf{B}$ projection on ($X$,$Y$) plane.

The polarization maps calculated for the disc models with $\dot{M}=10^{-8}$ and $\dot{M}=10^{-6}$~M$_{\sun}$~yr$^{-1}$ are very similar to those shown in Fig.~\ref{Fig:8}.

It should be noted that the disc is optically thin at 237~GHz for the considered disc models and dust opacity. One can expect that in the case of optically thick disc the polarization maps can be different from those presented in this section as the continuum emission will trace different regions of the disc and, thus, different magnetic field geometry.

\section{Conclusions and discussion}
\label{Sec:Summary}

We investigated the fossil magnetic field in the accretion and protoplanetary discs of young stellar objects. The accretion disc model described in DK14 includes Shakura and Sunyaev equations, the induction equation with Ohmic diffusion and magnetic ambipolar diffusion, the equations of ionization-recombination balance. We take into account ionization by cosmic rays, X-rays, radiative recombinations, recombinations onto dust grains, and thermal ionization.

We investigated the influence of the magnetic field buoyancy and the Hall effect on the magnetic field in the disk, which were not taken into account in our previous study (DK14). The stationary solution of the induction equation has the form, in which the buoyancy represents the additional mechanism of the magnetic flux loss. Such a modification allows us to investigate the fossil magnetic field both in the regions of the effective generation, and in the `dead' zones. We also take into account non-linearity of magnetic ambipolar diffusion (MAD) in the modified model. 

Our model is the useful tool for the investigation of dynamics of accretion and protoplanetary discs. In contrast to the other investigations, where the magnetic field strength and/or geometry are a priori determined, in our model the strength and geometry of the magnetic field in the discs are calculated. The predictions of the model are in agreement with observational constraints, as discussed below.

The radial and azimuthal magnetic field components are zero at $z=0$ due to the equatorial symmetry of the disc. In recent shearing-box MHD simulations, the generation of the strong azimuthal magnetic field in the midplane due to the Hall effect was found \citep{bai14, lesur14, bai16}. This result is caused by the peculiarities of the shearing-box approximation which does not allow to distinguish between the odd and even symmetries of the magnetic field. Moreover, the artificial limitation of the diffusivity in the region of the ionization fraction minimum was used in these calculations, which leads to the overestimation of the Hall effect efficiency.

Our calculations show that the buoyancy constraints the toroidal magnetic field generation. The strength of $B_{\varphi}$ is comparable with the vertical magnetic field strength in the inner region of the disc, $r\lesssim1$~au at $z\sim 0.5$~H. This result confirms simple estimates of DK14. The fossil magnetic field strength is nearly equal to the stellar magnetic field at the inner edge of the disc.  In the outer region, the non-linear magnetic ambipolar diffusion and buoyancy constraint the growth of the toroidal magnetic field, so that the toroidal magnetic field $B_{\varphi}$ remains comparable with $B_z$.

The Hall effect leads to the transformation of the azimuthal magnetic field to the radial one, and vice versa. The radial magnetic field becomes comparable with the azimuthal and vertical ones due to the Hall effect in the regions where electrons are magnetized. This happens near the borders of the `dead' zone. Thus, the magnetic field gains the quasi-radial geometry, $B_r \sim B_z$, in these regions at $z\sim 0.5$~H. The quasi-radial magnetic field promotes the generation of centrifugal wind \citep{blandford82}. DK14 have shown that magnetic ambipolar diffusion may prevent the generation of the quasi-radial magnetic field. On the basis of our new results, we conclude that the Hall effect is an important factor determining the possibility of the centrifugal wind launching in the accretion discs of young stars.

We calculated the geometry and strength of the fossil magnetic field in discs with different accretion rates. We refer to these cases as the evolutionary sequence from the protostellar to the protoplanetary discs. The protostellar discs are characterized by the higher accretion rate, $\dot{M}=10^{-6}\,\rm{M}_{\odot}\,\rm{yr}^{-1}$. The case with $\dot{M}=10^{-7}\,\rm{M}_{\odot}\,\rm{yr}^{-1}$ corresponds to the accretion disc of a young T~Tauri star. In the case of the lower accretion rate $\dot{M}=10^{-8}\,\rm{M}_{\odot}\,\rm{yr^{-1}}$, the disc is considered to be protoplanetary. 

Calculations show that the geometry of the fossil magnetic field does not depend significantly on the accretion rate. In all cases, the magnetic field remains quasi-azimuthal in the inner region, quasi-vertical inside the `dead' zone, quasi-radial or quasi-azimuthal in the outer regions of the disc. The `dead' zone achieves its maximum size in the protostellar disc. The extent of the `dead' zone decreases with the decrease of the accretion rate. The inner and outer boundaries of the `dead' zone move closer to the star. The strength of the fossil magnetic field goes down at any given distance during the evolution. For example, the vertical magnetic field strength falls down from $\sim 0.1$~Gs ($\dot{M}=10^{-6}\,\rm{M}_{\odot}\,\rm{yr}^{-1}$, protostellar disc) to $\sim 0.01$~Gs ($\dot{M}=10^{-8}\,\rm{M}_{\odot}\,\rm{yr^{-1}}$, protoplanetary disc) at $r=3$~au.	These values coincide with the remnant magnetic field strength inferred from the meteorites magnetization measurements \citep{levy78, fu14}.

We constructed the synthetic maps of the dust emission polarized due to the alignment of the dust grains with the magnetic field direction. The combination of the quasi-azimuthal and the quasi-radial magnetic field geometries in the inner disc appear as the spiral magnetic field structure in the face-on disc map. The synthetic polarization map for the face-on disc shows that it will be possible to spatially resolve the `dead' zones in nearby accretion discs in observations with the instruments like The Atacama Large Millimeter/Submillimeter Array (\textsc{ALMA}).  The `dead' zones will appear as a hole-like central regions where the value of polarization degree is small compared to those in the adjacent parts of the disc. Thus, the observations of the polarized emission can be a useful tool for the investigation of the planet formation region properties in the protoplanetary discs. Our synthetic maps are more detailed than the current observational data. In recent observations, it has been found that the magnetic field may be toroidal \citep{scox15} or the combination of toroidal and poloidal \citep{stephens14} in the protostellar and protoplanetary discs. We predict that we will see the different types of the magnetic field geometries in the different parts of the discs. For example, the quasi-radial magnetic field still has not been observed in the outer regions of protoplanetary discs. We hope that future polarization measurements will confirm our predictions.
 
We stress out that the radiative transfer calculations presented in this work are not aimed to reproduce the observations and to perform the detailed simulations of disc observations because of the simplicity of our disc model. The model does not account the dust growth, settling or radial drift that can affect the dust size and spatial distribution and, thus, the polarization of the disc emission \citep[see e.g.][]{Cho_etal2007}. The detailed observations of the continuum emission polarization may be useful to investigate the dust dynamics and protoplanet formation process. Moreover, we do not consider the cases when the disc is embedded in a dense envelope.

In the present work we haven't considered UV radiation from the central star and the role of dust particles charge. The stellar UV ionizes only thin surface layers of the disc. Our conclusions about the magnetic field inside the disc will not change if UV radiation will be taken into account.  Examinations of more complex ionization model including different dust grain charges will be addressed to the future papers. We neglected the effect of the charged dust grains on the anisotropy of the conductivity tensor. Charged dust grains with mass $m_{\rm{g}}$ and charge $e$ can produce the anisotropy of the conductivity only in case when their magnetization parameter
\begin{equation}
	\beta_{\rm{g}} = \frac{eB}{m_{\rm{g}}c}\frac{1}{\nu_{\rm{gn}}}\label{Eq:beta_g1}
\end{equation}
is more than unity. In (\ref{Eq:beta_g1}), $\nu_{\rm{gn}}=\langle\sigma v\rangle_{\rm{gn}}n_{\rm{n}}$ is the frequency of collisions of the grains with the neutrals, $\langle\sigma v\rangle_{\rm{gn}}\simeq \pi a_{\rm{d}}\sqrt{\frac{8kT}{\pi m_{\rm{n}}}}$ is the corresponding coefficient of the momentum transfer, $m_{\rm{n}}$ is the mass of the neutral particle. Assuming that the density of the dust grains $\rho_{\rm{g}}=2\,\rm{g}\,\rm{cm}^{-3}$, we derive from (\ref{Eq:beta_g1}) with the typical parameters at $1$~au in our model (see Figs.~\ref{Fig:1}, \ref{Fig:2}, \ref{Fig:4})
\begin{equation}
	\beta_{\rm{g}} = 7\times 10^{-17}\left(\frac{B}{0.1\,\rm{Gs}}\right)\left(\frac{a_{\rm{d}}}{0.1\,\mu\rm{m}}\right)^{-5}\left(\frac{T}{800\,\rm{K}}\right)^{-\frac{1}{2}}\left(\frac{n_n}{10^{14}\,\rm{cm}^{-3}}\right)^{-1}.\label{Eq:beta_g2}
\end{equation}
Therefore, the dust particles with sizes $a_{\rm{d}}\geq 0.1\,\mu$m (considered in our calculations) are too large to be magnetized and contribute to the anisotropy of conductivity. Expression (\ref{Eq:beta_g2}) shows that the dust grains with $a_{\rm{d}}<6\times 10^{-3}\,\mu$m will be magnetized with the adopted values of density, temperature, and magnetic field strength. It should be noted that the radius of the dust grains at $T=800$~K is 15 times smaller than the initial radius $a_{\rm{d}}=0.1\mu$m at $T<150$~K due to the evaporation of ices in our calculations. Such dust grains are also non-magnetized according to (\ref{Eq:beta_g2}). The evaporation of the dust grains in the inner regions of the discs should be taken into account in the conductivity calculations. The evaporation of the dust grains causes the reduction of the recombination rate which leads to the growth of the ionization fraction.

Turbulence in our model is described in terms of \citet{ss73} $\alpha$ approximation. In order to determine the turbulent diffusivity and its influence on the magnetic field correctly, it is required to study MHD turbulence properties by solving the Reynolds equations \citep[e.g., ][]{ruzmaikin88}. Turbulence leads to the generation of the small scale magnetic field out of the large-scale one. The question whether the turbulence can transport the large-scale magnetic field requires further analysis. This complex issue is beyond the scope of current work. Our further investigations will focus on the dynamical influence of the magnetic field on the structure of the accretion discs.

\section*{Acknowledgements}
We thank anonymous referee for his/her useful comments and suggestions that help us to present our results more clearly.
This work is supported by Russian Science Foundation (project 15-12-10017). This research used \textsc{aplpy} package hosted at \href{http://aplpy.github.com}{http://aplpy.github.com}. All figures in this paper were plotted with the \textsc{matplotlib} package \citep{Hunter:2007}.

\bibliographystyle{mnras}
\bibliography{khaibrakhmanov}

\bsp	
\label{lastpage}

\end{document}